\documentclass[a4paper,11pt]{article}
\usepackage{pos,slashed,dsfont}

\newcommand{\Bielefeld}{Fakult\"at f\"ur Physik, Universit\"at Bielefeld,
  D-33615 Bielefeld, Germany.}
\newcommand{\Budapest}{E\"otv\"os Lor\'and University, P\'azm\'any P\'eter\ s\'et\'any\ 1/A, H-1117, Budapest
  Hungary.}
\newcommand{\Lendulet}{Institute for Nuclear Research, Bem t\'er 18/c, H-4026 Debrecen, Hungary.}

\newcommand{\be}{\begin{equation}}
\newcommand{\ee}{\end{equation}}
\newcommand{\dd}{\textmd{d}}

\newcommand{\Z}{\mathcal{Z}}

\newcommand{\expv}[1]{\left \langle #1 \right \rangle}

\title{Role of inhomogeneities in the flattening of the
quantum effective potential}

\author[a]{Gergely Endr\H{o}di}
\author[b,c]{Tam\'as G. Kov\'acs}
\author*[a]{Gergely Mark\'o}

\affiliation[a]{\Bielefeld}
\affiliation[b]{\Budapest}
\affiliation[c]{\Lendulet}

\emailAdd{endrodi@physik.uni-bielefeld.de}
\emailAdd{tamas.gyorgy.kovacs@ttk.elte.hu}
\emailAdd{gmarko@physik.uni-bielefeld.de}

\abstract{We investigate the role of inhomogeneous field configurations in systems with a spontaneously broken continuous global symmetry. Spontaneous breaking is usually defined as a specific double limit, first infinite volume at finite explicit breaking sources, which are then extrapolated to zero. We consider a different approach in which the order parameter is constrained under the path integral, which we simulate using lattice Monte Carlo techniques. In this way we access the flat region of the effective potential and we show that inhomogeneous configurations are dominant there. We topologically classify the important configurations and measure the excess energy stored in the inhomogeneities allowing for the definition of a generalized differential surface tension. We show that this contribution becomes negligible at large volumes restoring the flatness of the effective potential in the thermodynamic limit.}

\FullConference{%
 The 38th International Symposium on Lattice Field Theory, LATTICE2021
  26th-30th July, 2021
  Zoom/Gather@Massachusetts Institute of Technology
}

\begin{document}
\maketitle
\section{Introduction}
One of the most ubiquitous phenomena in physics is the spontaneous breaking of symmetries. Examples range from condensed matter to high energy physics and in classical as well as quantum systems. More directly in quantum field theory prime examples are the Higgs mechanism, the chiral symmetry breaking of strongly interacting matter or even the Peccei-Quinn mechanism in the theory of axions.

Furthermore, in all of the above examples in field theories, inhomogeneous order parameters were also considered. In the Higgs sector this means impurities in the condensate \cite{Hosotani:1982ii,Kibble:1997yn,GarciaBellido:2002aj,Giudice:2010zb}, while for axions inhomogeneous miniclusters may exist \cite{Marsh:2015xka}. In QCD and in lower dimensional similar models, the very dense strongly interacting matter is expected to form inhomogeneous chiral-spirals \cite{Buballa:2014tba,Kojo:2009ha,Lenz:2020bxk,Lenz:2021kzo,Lenz:2021vdz}.

In this work we discuss inhomogeneities, which emerge due to the coexistence of phases corresponding to different orientations of the order parameter. 
This coexistence is responsible for the flatness of the effective potential, which is well described by the formation of bubbles of different phases when the broken symmetry is {\it discrete}. The inhomogeneity of the local order parameter allows the average order parameter to change continuously between the discrete set of homogeneous phases, with the excess energy due to the inhomogeneity being stored in the bubble walls. The flatness of the effective potential in the thermodynamic limit is due to the fact that the surface energy becomes negligible in the energy density for large volumes. In the case of a {\it continuous} symmetry being spontaneously broken, bubble walls do not form since the local order parameter can change continuously. In this case we expect spin-wave like structures within the coexistence region of the effective potential \cite{Ringwald:1989dz,Mukaida:1996sr}. In what follows we demonstrate this in a simple model, uncovering finer details of this mechanism as well as introducing a well quantifiable generalization of the bubble wall surface tension. More details of our results can be found in Ref.~\cite{Endrodi:2021kur}.

\section{The model and the method}
We discuss specifically the three dimensional $O(2)$ symmetric $\phi^4$ model defined by the Lagrangian density
\begin{equation}
\label{eq:lagrange}
\mathcal{L}(x)= \frac{m^2}{2}\sum_a\phi_a^2(x) +\frac{g}{24}\Big[\sum_a\phi_a^2(x)\Big]^2\!+\,\frac{1}{2}\sum_{\mu,a}[\partial_\mu \phi_a(x)]^2 \,,
\end{equation}
with $a=0,1$ denoting the two components of the scalar field. We investigate the parameter region where $m^2<0$ and $g>0$, so the model is in the spontaneously broken phase. Shifting \eqref{eq:lagrange} by a spatially constant, linear explicit symmetry breaking term $h\phi_0(x)$ in the $a=0$ direction (without loss of generality) the effective potential of the model has a well defined minimum. The textbook definition of the value of the order parameter in the spontaneously broken case (denoted by $\bar\phi_{\rm min}$) is then given by the following limiting procedure. $V^{-1}\langle\sum_x\phi_0(x)\rangle_h$, the expectation value of the volume average of the field at some value of $h$ is evaluated, slowly removing the explicit breaking, strictly in the thermodynamic limit. Notice that the removal of $h$ is also driving the system into a first order type of criticality as the direction of the order parameter depends entirely on the direction of $h$. Therefore at $h=0$ we refer to the values of the order parameter $\bar\phi<\bar\phi_{\rm min}$ regardless of direction as being inside the coexistence disk. Our goal is to explore the coexistence disk as well as to understand and describe the inhomogeneities leading to the flatness of the effective potential in this region.

The partition function of the system in the presence of the explicit breaking term can be written as
\be
Z_h=\int [\dd\phi_a(x)] \exp\left\{-\int \!\dd^3 x \left[ \mathcal{L}(x)-h\,\phi_0(x)\right]\right\}\,,
\label{eq:Z}
\ee
which we refer to as the {\it canonical} partition function and objects derived from it as being in the canonical formulation (the corresponding expectation values are denoted by $\expv{.}_h$). The effective potential is then defined as the Legendre transform of $\log Z_h$ with respect to $h$,
\be
\Gamma(\bar\phi) \equiv \underset{h}{\rm sup} \left(h\cdot \bar\phi - \frac{1}{V} \log Z_{h}\right)\,.
\label{eq:Gammadef}
\ee
As the {\it microcanonical} counterpart, one can define a constrained partition function, where the volume average of the field is restricted to a certain value ($\bar\phi$ in one direction and vanishing in the other in line with the choice of $h$ in the canonical formulation)
\be
\Z_{\bar\phi}\!=\! \int [\dd\phi_a(x)]\, e^{-\int\!\dd^3x\, \mathcal{L}(x)}\,\delta^{(2)}\!\left(\frac{1}{V}\!\int\!\dd^3 x\, \phi_b(x)-\bar\phi \,\delta_{b0}\!\right)\,.
\ee
In the microcanonical formulation one can define the constraint potential $\Omega(\bar\phi)=-\log\Z_{\bar\phi}/V$ as well as expectation values according to $\Z_{\bar\phi}$ (denoted by $\expv{.}_{\bar\phi}$). One can show that in the thermodynamic limit $\Omega(\bar\phi) = \Gamma(\bar\phi)$ \cite{ORaifeartaigh:1986axd,Kuti:1987bs,Heller:1983xg}. Consequently, their respective $\bar\phi$ derivatives also coincide in and only in the infinite volume limit.

We carry out Monte Carlo lattice simulations according to the constrained partition function on $V=L^3$ lattices (L=\{40, 60, 80\}), using the model parameters $m^2=-15.143$ and $g=102.857$. The simulated range of $\bar\phi$ is from $0$ to $\bar\phi \gtrapprox \bar\phi_{\rm min}$. The generation of fixed $\bar\phi$ configurations is done using a hybrid Monte Carlo algorithm based on \cite{Fodor:2007fn}.

\section{Results}

\begin{figure}[t]
\begin{center}\includegraphics[width=0.7\textwidth]{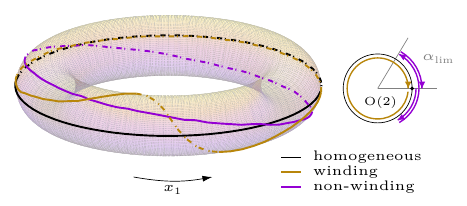}
\end{center}\caption{\label{fig:torus} Examples for the dominant configurations of the constrained path integral in the $\bar\phi<\bar\phi_{\rm min}$ region. The torus represents the $O(2)$ space and the $x_1$ coordinate and the curves depict $\Sigma(x_1)$ in the $w=1$ and $w=0$ cases. All lines are solid (dashed) when they are in front of (behind) the torus. The example configurations were generated at $\bar\phi\approx0.5$ on a $V=80^3$ lattice.}
\end{figure}

We first examine the configurations appearing in the fixed $\bar\phi$ Markov-chains. We find that configurations giving the dominant contribution to the partition function (and to observables) are inhomogeneous even after smearing out ultraviolet (UV) fluctuations. The inhomogeneities are one dimensional, therefore, without loss of generality, we can always choose the $x_1$ spatial coordinate to point in that direction. In order to characterize the configurations based on their behaviour in the $x_1$ direction we define $x_1$-slices of the field by averaging in the other two directions to suppress UV fluctuations:
\be
\label{eq:slicedef}
\Sigma_a(x_1)\!=\!L^{-2}\int \!\dd^3 y \,\phi_a(y) \,\delta(y_1-x_1)\,.
\ee
We find that the magnitude of $\Sigma(x_1)$ is around $\bar\phi_{\rm min}$ everywhere, however its direction changes as a function of $x_1$. Specifically, an integer winding number $w$ can be assigned to each configuration based on how many times the $\mathrm{O}(2)$ group is mapped to the circle corresponding to the periodic $x_1$ coordinate. Finally, we find that the dominant contributions are coming from configurations with $w=1$ or $w=0$. Hence, we refer to the two sectors as {\it winding} or {\it non-winding} cases. An example pair of configurations is plotted next to a homogeneous configuration in Fig.\ref{fig:torus}, highlighting the topological aspects. One can see that only the non-winding configuration can be continuously deformed into a homogeneous one. From a simulational viewpoint, we note that due to the topological nature of the configurations, tunnelings between the two sectors are rare even when the relative action differences are small. Later on we see that this is a relevant effect only in a small window in $\bar\phi$ and that instead it is sufficient to measure observables in separate sectors (denoted by a $w$ index). 

The two configuration types can be approximated using a common (up to the winding number) ansatz motivated by classical solutions of the equations of motion (in the presence of the constraint). The ansatz assumes a constant length of the field and only the $O(2)$ angle depends on $x_1$.
\begin{subequations}
\label{eq:ansatz}
\begin{align}
\phi^{(w)}(x_1) &= \bar\phi_{\rm min}\cdot (\cos\alpha_w(x_1),\sin\alpha_w(x_1) )^\top\,,\\
\alpha_w(x_1) &= \frac{2\pi w x_1}{L} + \alpha_{\rm lim} \sin\left(\frac{2\pi x_1}{L}\right)\,.
\end{align}
\end{subequations}

\begin{figure}
\begin{center}\includegraphics[width=0.7\textwidth]{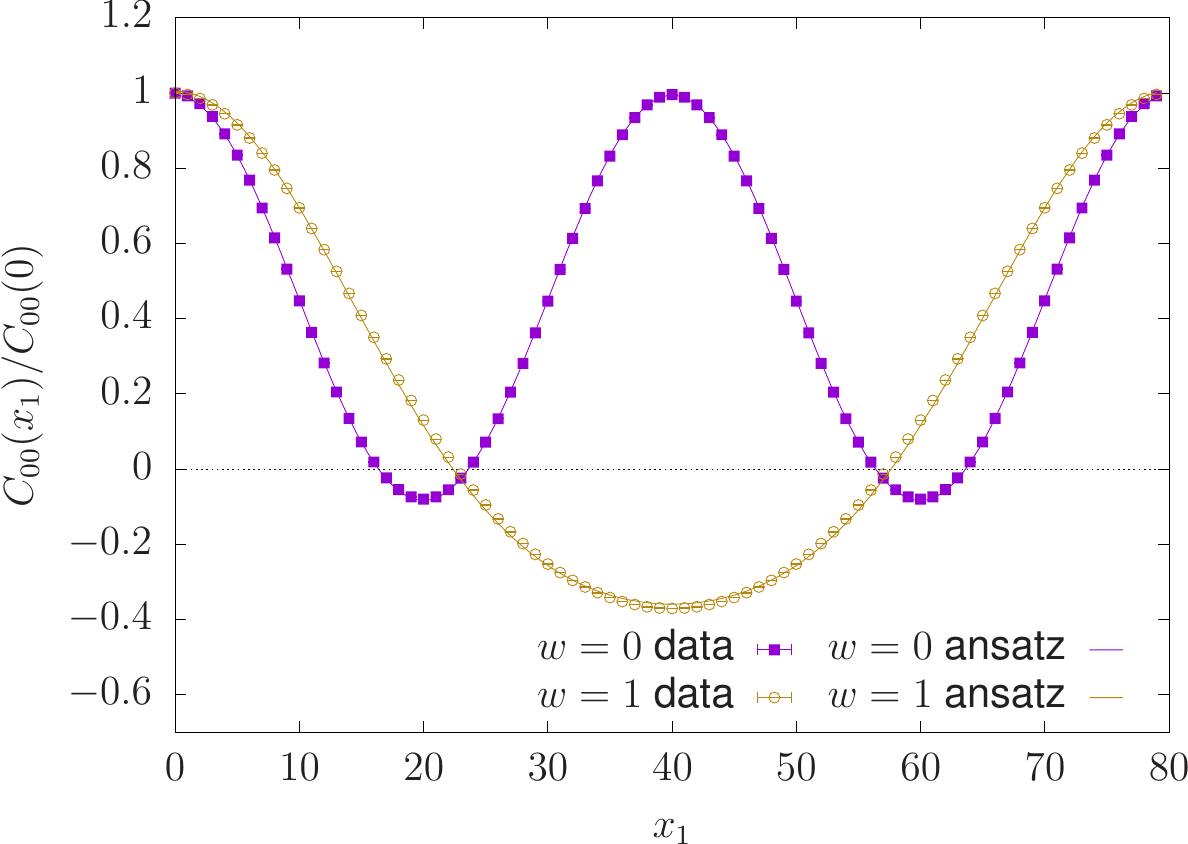}
\end{center}\caption{\label{fig:corrs} Slice correlators as measured in the winding and non-winding sectors for $V=80^3$, at $\bar\phi\approx0.25$. The respective lines following closely the two datasets are the correlators calculated from the classical ansatz, where the parameter $\alpha_{\rm lim}$ was optimized in a least squares fashion.}
\end{figure}

Using the classical ansatz not only the configurations can be fitted well, but certain observables can also be very well approximated. In Fig.~\ref{fig:corrs} we show the $00$ component of the slice correlator matrix
\be
C_{ab}(x_1)=\expv{\Sigma_a(x_1)\Sigma_b(0)}_{\bar\phi}
\ee
in the two topological sectors compared to the convolutions of the respective ansatz functions. Notice that for the correlators, inhomogeneities survive the ensemble averaging, which one would not be able to see by looking at e.g. the 1-point function at a certian value of $x_1$.

\begin{figure}[t]
\begin{center}\includegraphics[width=0.7\textwidth]{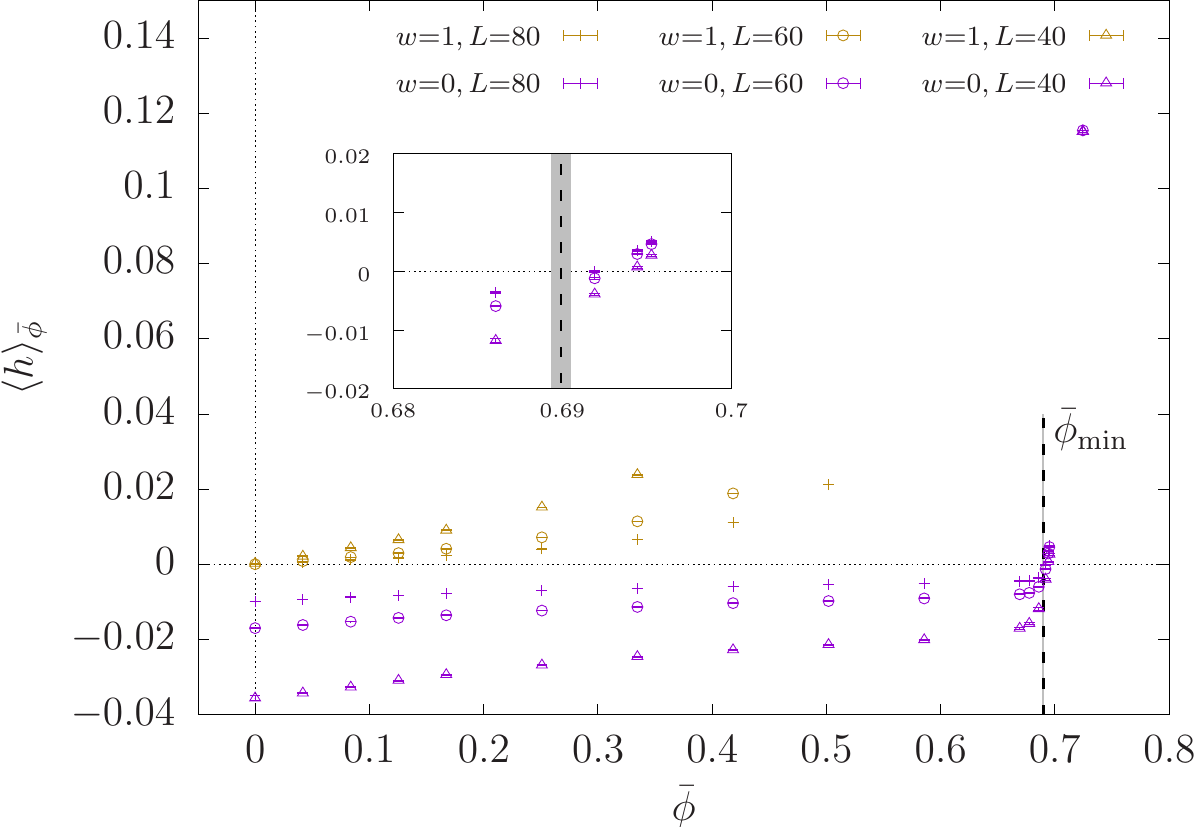}
\end{center}\caption{\label{fig:h_of_phi} The magnetic field expectation value $\langle h\rangle_{\bar\phi,w}$ as a function of $\bar\phi$. The gold (purple) points correspond to measurements in the $w=1(0)$ sector. Also shown is the thermodynamic limit extrapolation of $\bar\phi_{\rm min}$ obtained from the intersect of the results with $\langle h\rangle_{\bar\phi}=0$. The region in the vicinity of $\bar\phi_{\rm min}$ is highlighted by the inset.}
\end{figure}

We also want to investigate how exactly the inhomogeneities contribute to the flattening of the effective potential. In the microcanonical formulation we cannot access this quantity directly, instead we can reconstruct the constraint potential, $\Omega(\bar\phi)$. We use the integral method, by measuring
\be
\label{eq:hdef}
\expv{h}_{\bar\phi}
\equiv \frac{\partial \Omega(\bar\phi)}{\partial \bar\phi}
= m^2 \bar\phi + \frac{g}{6V} \expv{\int\!\dd^3x \sum_{a}\phi_a^2(x)\phi_0(x)}_{\!\bar\phi}\,,
\ee
and integrating with respect to $\bar\phi$. As mentioned earlier, we measure $\expv{h}_{\bar\phi,w}$ in the two sectors separately, and therefore the constrained potential in the respective sectors is
\be
\label{eq:omegaint}
\Omega_{w}(\bar\phi) = \int_0^{\bar\phi} d\bar\phi' \,\expv{h}_{\bar\phi',w} + c_{w}\,,
\ee
where $c_w$ are integration constants to be set later. The results for $\expv{h}_{\bar\phi,w}$ are shown in Fig.~\ref{fig:h_of_phi}, revealing how the magnetic field approaches zero in the thermodynamic limit from opposite directions for $w=0$ and $w=1$. Notice that the vanishing point of $\expv{h}_{\bar\phi,0}$ is the finite volume estimate for the edge of the coexistence disk, which can be interpolated with a good precision. The extrapolation to the thermodynamic limit gives $\bar\phi_{\rm min}=0.6899(6)$. 

\begin{figure}
\begin{center}\includegraphics[width=0.7\textwidth]{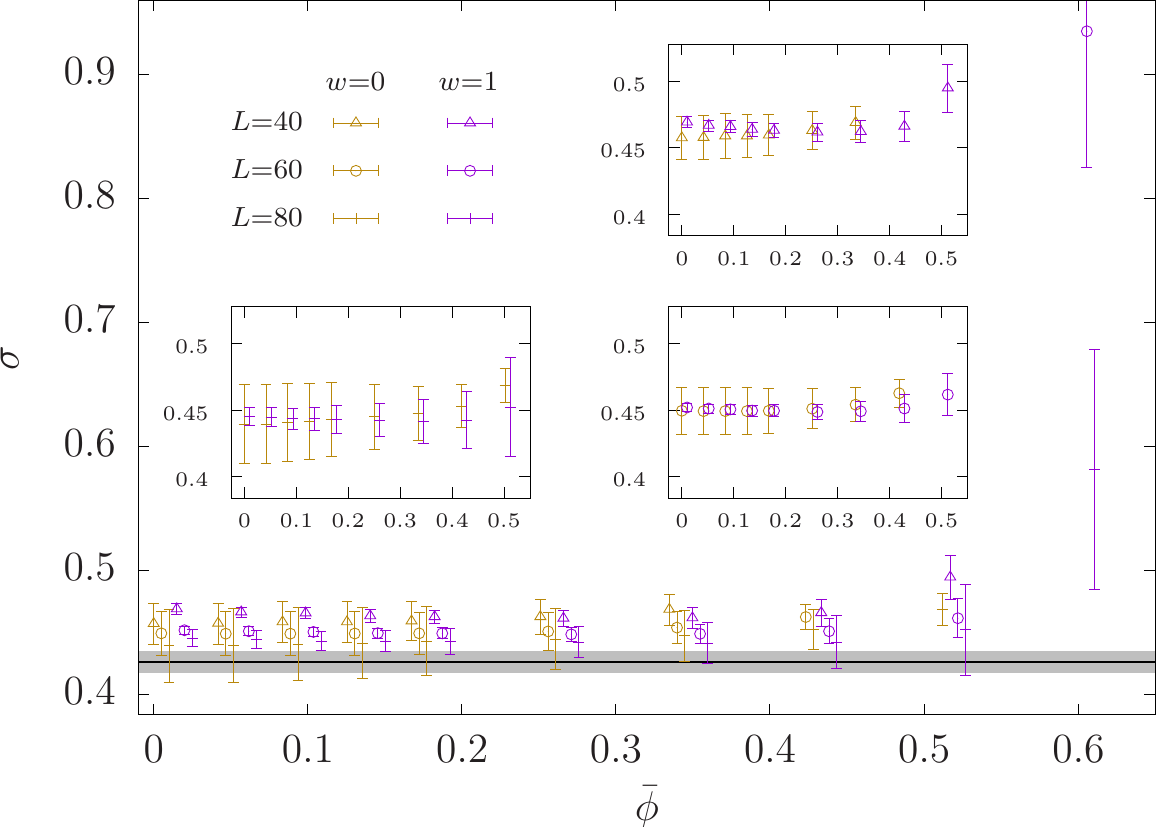}
\end{center}\caption{\label{fig:surftens} The differential surface tension as defined in Eq.~\eqref{eq:surfacetension_formula} for three different volumes in both relevant topological sectors, together with the infinite volume limit, $\sigma=0.427(8)$ (gray band). The insets show each volume separately to reveal more clearly the independence of $\sigma$ of the topological sector. Some of the points are shifted horizontally for better visibility.}
\end{figure}

For the full reconstruction of the constraint potential we need the integration constants $c_{0}$ and $c_{1}$. The non-winding case can be set simply since this sector connects continuously to the edge of the coexistence disk. Hence we require that $\Omega(\bar\phi_{\rm min})_0=0$ which sets the value of $c_0$.
In the winding case we use that a) the dominant configuration type changes at some intermediate $\bar\phi_{\rm c}$, b) the excess energy compared to $\Omega(\bar\phi_{\rm min})$ stems from the inhomogeneity of the configurations and can be characterized by a generalization of the surface tension known from bubble formation. Statement b) is made formal by introducing the kinetic energy of the sliced fields (again to suppress UV fluctuations)
\be
E_\Sigma(\bar\phi) = \frac{1}{2L}\sum_{a,x_1}\expv{[\partial_{x_1} \Sigma_a(x_1)]^2}_{\bar\phi}\,.
\ee
Then we parametrize the excess energy for a $\bar\phi$ where inhomogeneities are already dominant as
\be
\Omega(\bar\phi)= \sigma \cdot \frac{E_\Sigma(\bar\phi) -E_\Sigma(\bar\phi_{\rm min})}{\bar\phi_{\rm min}^2}\,,
\label{eq:surfacetension_formula}
\ee
where we used that we set $\Omega(\bar\phi_{\rm min})=0$, hence no subtraction is necessary on the left hand side. Statement b) can then be reformulated as the independence of $\sigma$ of both $\bar\phi$ and $w$. Notice that for a discrete symmetry system with a domain wall of characteristic width $\Delta$ separating phases with $+\bar\phi_{\rm min}$ and $-\bar\phi_{\rm min}$, the same formula gives $V\Omega = \sigma \cdot L^2$, where $\sigma\propto 1/\Delta$ and the proportionality factor depends on the precise profile of the wall. Therefore $\sigma$ is an appropriate generalization of the surface tension and we refer to it as {\it differential} surface tension. By confirming that $\sigma$ is indeed constant (see Fig.~\ref{fig:surftens})\footnote{The errors on the points close to $\bar\phi_{\rm min}$ are blowing up , since $\sigma$ is obtained as the numerical evaluation of a $0/0$ type limit as the edge of the coexistence disk is approached.} we see that the transition from one dominant topological sector to the other happens where $E_{\Sigma,  0}=E_{\Sigma,1}$ and at that point $\Omega_0=\Omega_1$ as well. This allows us to determine $\bar\phi_{\rm c} = 0.2818(2)$ (already extrapolated to the thermodynamic limit) as well as the integration constant $c_1$ for the reconstruction of the constraint potential in the winding sector.

\begin{figure}
\begin{center}\includegraphics[width=0.7\textwidth]{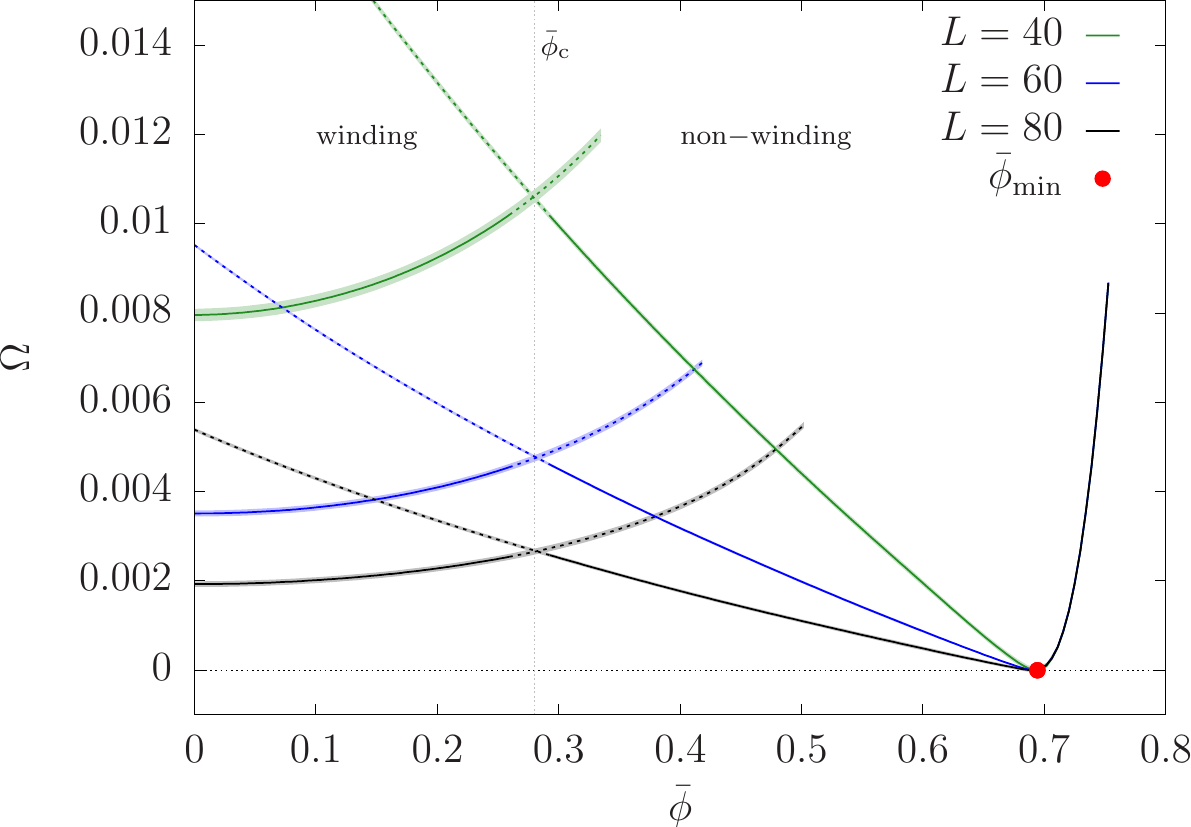}
\end{center}\caption{\label{fig:pots} The constraint potential for different volumes. Regions are labeled based on which configuration type is dominant there, and correspondingly lines are solid (dashed) in the regions where they are (sub-) dominant. The potential is consistent everywhere with a flat thermodynamic limit, assuming an $L^{-2}$ scaling.}
\end{figure}

Finally we reconstruct the full constraint potential, except in the immediate vicinity of $\bar\phi_{\rm c}$. We can always use the $\Omega_w$ of the dominant sector as the other configurations are exponentially suppressed in terms of relative probabilities. The results are plotted in Fig.~\ref{fig:pots}

\section{Discussion}
We studied the three dimensional $O(2)$ symmetric scalar $\phi^4$ model using constrained Monte Carlo simulations to discuss some general features of spontaneously broken global symmetries. We found that within the coexistence disk the constrained path integral and the corresponding potential $\Omega(\bar\phi)$ is dominated by inhomogeneous field configurations. The macroscopic inhomogeneities can be described by approximately constant length spin-waves along a certain axis. These configurations can be assigned into topological sectors based on their winding number $w$. While for values close to $\bar\phi=0$ the winding configurations are more important, at an intermediate value $\bar\phi_{\rm c}$, the $w=0$ configurations become much more likely through a sharp transition. The other $w$ sectors are suppressed everywhere.

We introduced the differential surface tension $\sigma$ as a generalization of the usual excess energy stored in bubble walls for discrete symmetry systems. In continuous symmetry systems bubble walls do not form since the local order parameter can change continuously. Therefore the excess energy stored in the inhomogeneity can be captured locally but nevertheless can be characterized by a global quantity. We show that our definition is independent of $\bar\phi$ and of $w$ and allows us to reconstruct practically the full constraint potential. Since the energy stored in the inhomogeneities is proportional to the linear length of the system, we show that the constraint potential indeed tends to the flat effective potential in the thermodynamic limit.\\

\noindent
{\bf Acknowledgments} 
This research was funded by the DFG (Emmy Noether Programme EN 1064/2-1 and the Collaborative Research Center CRC-TR 211 ``Strong-interaction
matter under extreme conditions'' -- project number
315477589 - TRR 211). 

\bibliography{flatproc}
\bibliographystyle{utphys}

\end{document}